\documentclass[sigconf,nonacm]{acmart}
\usepackage{mathtools} % This package enhances math typesetting  
\usepackage{array}
\usepackage{multirow}

\AtBeginDocument{%
  }

\setcopyright{acmlicensed}
\copyrightyear{2026}
\acmYear{2026}
\acmDOI{XXXXXXX.XXXXXXX}

\acmConference[RecSys '26 Accepted paper]{ACM Conference on Recommender Systems}
  {September 28--October 2, 2026}{Minneapolis, Minnesota, USA}
\acmISBN{978-1-4503-XXXX-X/18/06}

\begin{document}

\title{MESH: Scaling Up Retrieval with Heterogeneous Content Unification}
%Scaling Unified Retrieval for Heterogeneous Ecosystems with Structured Tokenization
%MP-Mixer
%MESH

\author{Jiaxing Qu}
\email{jqu@pinterest.com}
\affiliation{%
  \institution{Pinterest Inc.}
  \city{San Francisco}
  \state{CA}
  \country{USA}
}

\author{Yilin Chen}
\email{yilinchen@pinterest.com}
\affiliation{%
  \institution{Pinterest Inc.}
  \city{San Francisco}
  \state{CA}
  \country{USA}
}

\author{Junpeng Hou}
\email{jhou@pinterest.com}
\affiliation{%
  \institution{Pinterest Inc.}
  \city{San Francisco}
  \state{CA}
  \country{USA}
}

\author{Jinfeng Rao}
\email{marquisrao@pinterest.com}
\affiliation{%
  \institution{Pinterest Inc.}
  \city{San Francisco}
  \state{CA}
  \country{USA}
}

\author{Olafur Gudmundsson}
\email{ogudmundsson@pinterest.com}
\affiliation{%
  \institution{Pinterest Inc.}
  \city{San Francisco}
  \state{CA}
  \country{USA}
}

\author{Sai Xiao}
\email{sxiao@pinterest.com}
\affiliation{%
  \institution{Pinterest Inc.}
  \city{San Francisco}
  \state{CA}
  \country{USA}
}

\author{Huizhong Duan}
\email{hduan@pinterest.com}
\affiliation{%
  \institution{Pinterest Inc.}
  \city{San Francisco}
  \state{CA}
  \country{USA}
}

\renewcommand{\shortauthors}{Jiaxing Qu et al.}

\begin{abstract}
Optimizing large-scale retrieval hinges on the ability to efficiently surface candidates across diverse content tiers. However, to capture segments such as fresh and long-tail content, modern systems typically resort to a fragmented ``zoo'' of specialized retrieval models. 
This operational complexity is attributed to a fundamental challenge in heterogeneous retrieval systems --- the Scaling Bias of Heterogeneity --- where model capacity gains do not apply equally across diverse 
content tiers.
%In this paper, we conduct extensive offline and online experimentation and report a key empirical finding: traditional two-tower models display a pronounced scaling divergence in heterogeneous recommendation ecosystems. Increasing model capacity consistently benefits high-frequency evergreen content, but we discover that sparse fresh and long-tail segments stagnate—a phenomenon we term \textit{The Scaling Bias of Heterogeneity}. 
To bridge this gap, we propose MESH as a unified retrieval scaling framework that mitigates this bias through a modularized architecture integrated with gated bias correction.
By partitioning the feature space into independent domains, MESH enforces a structural inductive bias that reduces interference between sparse-item signals and high-frequency engagement features.
This protected gradient path leads to improved scaling behavior for sparse content, empirically validated by a $14\times$ improvement in the power-law scaling exponent for fresh items.
In online evaluations on Pinterest's Related Pins platform — a billion-scale item-to-item recommendation system — these improvements translate into a +5.5\% lift in fresh-item repins, alongside with 55\% improvement in funnel efficiency and +0.46\% improvement in user retention.
Finally, our asynchronous serving strategy ensures production viability by delivering a 2.87$\times$ improvement in system throughput. 
Our findings suggest MESH as a promising paradigm for consolidating fragmented retrieval infrastructures into more scalable and ecosystem-aware backbones.
\end{abstract}

\thanks{Accepted -- ACM Conference on Recommender Systems
(RecSys 2026) Industry Track.}




%   \includegraphics[width=\textwidth]{closeup_exp.png}
%   \caption{Recommendation funnel and visual shopping journey in Closeup.}


\maketitle

\section{Introduction}
The success of large language models (LLMs) has sparked growing interest in discriminative scaling laws for recommendation systems, suggesting that performance can scale predictably with increases in model parameters and training data. 
To harness this potential, modern architectures \cite{zhang2025onetrans, zhu2025rankmixer, zhang2024wukong, chai2025longer} have championed various high-capacity neural backbones—ranging from factorized machines and mixture-of-experts (MoE) based mixers to massive transformer sequences — with the expectation that sparsity and cold-start challenges can be addressed through expanded model capacity.
However, unlike the uniform modality of text in NLP, the heterogeneous nature of recommendation signals raises a critical question: does a larger model benefit all content equally?

%However, we identify a systemic \textit{Scaling Bias of Heterogeneity} that undermines this flat scaling paradigm in large-scale industrial platforms like Pinterest. 

\begin{figure*}[t]
  \centering
  \includegraphics[width=\linewidth]{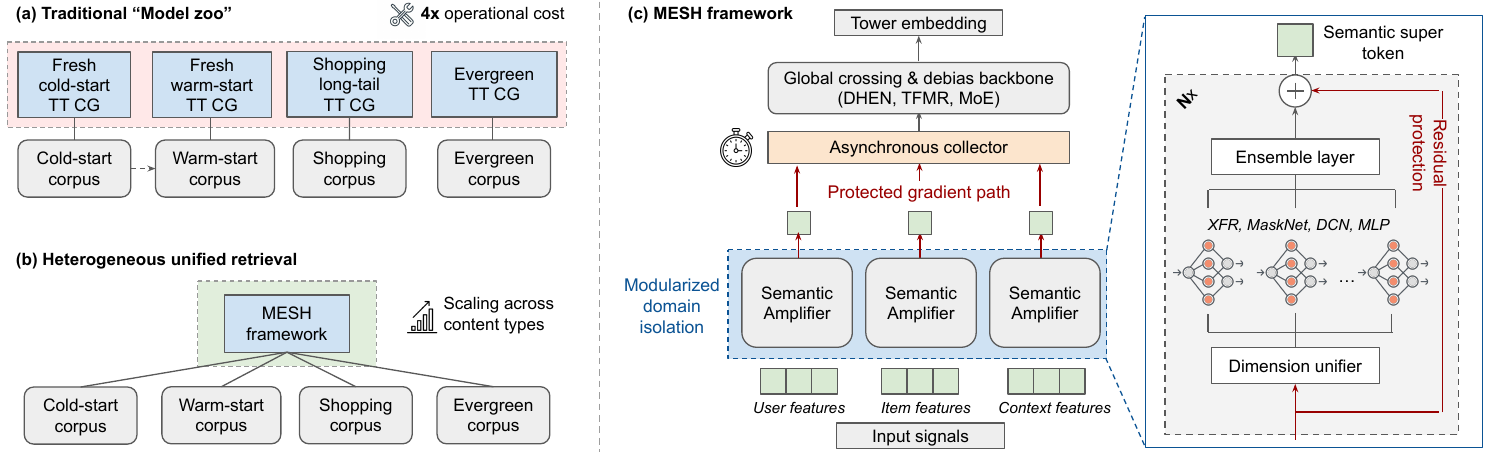}
    \caption{(a) The fragmented ``model zoo'': Traditional systems resort to a patchwork of retrieval channels for specialized content tiers (e.g. cold-start or long-tail). (b) A unified retrieval backbone that achieves heterogeneous scaling with significant operational cost savings. (c) The MESH Architecture. (Left) The unified retrieval pipeline featuring modularized domain isolation to protect the gradient paths of sparse entities, alongside an asynchronous collector for high-throughput serving. (Right) Detailed sub-tower internal structure, illustrating the recursive residual projection mechanism used to amplify raw signals into robust semantic super tokens. The schematic illustrates the architecture of user-side tower of two-tower model.}
    \label{fig:motivation}
\end{figure*}

To investigate this, we conduct a large-scale empirical study on Pinterest's billion-scale Related Pins platform, and observe a systematic scaling divergence: while increasing model capacity improves performance for high-frequency evergreen content, accuracy gains for sparse fresh and long-tail segments remain substantially smaller. We term this phenomenon as \textit{The Scaling Bias of Heterogeneity}. This divergence highlights an important challenge in unified retrieval systems: scaling compute primarily serves to refine the representations of popular items, creating a ``rich-get-richer'' feedback loop that marginalizes exploration of fresh and long-tail content.Consequently, new content 
remains trapped in a sub-optimal scaling regime, creating a gap between raw model capacity and overall ecosystem health.

%This creates an engagement bias that marginalizes fresh content, leading to a "rich-get-richer" feedback loop where scaling compute primarily serves to refine the representations of popular items, leaving new content trapped in a sub-optimal scaling regime with poor scaling exponent. This divergence represents a fundamental bottleneck, where scaling laws fail to translate into heterogeneous ecosystem health, primarily benefiting exploitation while marginalizing exploration.

%they prevent the system from achieving the full potential of scaling laws, as models are split across isolated, sub-scale components.

% This phenomenon is particularly acute in large-scale industrial platforms such as Pinterest's Related Pins — an item-to-item recommendation system that retrieves semantically related content from a billion-scale corpus across heterogeneous content tiers including evergreen, fresh, and long-tail items. 
To maintain segment-specific accuracy across tiers, practitioners often resort to a fragmented ecosystem of specialized two-tower (TT) models, such as maintaining dedicated candidate generators (CG) exclusively for cold-start retrieval (Figure \ref{fig:motivation}a).
%The prevalence of the ``model zoo'' is an empirical concession to this scaling bias.
Although these task-specific models offer temporary relief, this approach introduces substantial engineering overhead, additional infrastructure costs and operational burdens.
% Beyond operational burdens, partitioning parameters across isolated, sub-scale components prevents the system from capturing the full potential of global scaling laws. 
These inefficiencies motivate the need for a unified backbone capable of serving heterogeneous content tiers without sacrificing scalability or operational efficiency.

% To resolve these challenges, we propose MESH (\textbf{M}odularized \textbf{E}ffective \textbf{S}caling for \textbf{H}eterogeneity), a modularized semantic-tower architecture that bridges structural inductive bias with model scaling. 
% Unlike ``flat'' backbones that process features in a uniform space, MESH partitions heterogeneous signals into independent, domain-specific semantic amplifiers which leverages high-capacity hierarchical encoders and gated bias correction to stabilize feature interactions.
% By enforcing this structural decoupling, we effectively isolate the semantic manifolds of cold-start items from the extrinsic noise of engagement signals, providing a ``protected gradient path'' for sparse entities. 
% Our core contributions are three-fold:
To resolve these challenges, we propose MESH (\textbf{M}odularized \textbf{E}ffective \textbf{S}caling for \textbf{H}eterogeneity), a modularized retrieval architecture that partitions heterogeneous signals into independent domain-specific sub-towers before global interaction, providing a protected gradient path for sparse entities while retaining a unified and scalable backbone. Our core contributions are three-fold:
\begin{itemize}
    \item \textbf{Heterogeneous retrieval scaling}: Through a large-scale empirical study, we identify a systematic scaling divergence in flat unified retrieval architectures, where capacity gains are substantially stronger on evergreen content than on fresh and long-tail segments — a challenge we term the \textit{Scaling Bias of Heterogeneity}.
    \item \textbf{A modularized unified retrieval architecture}: We introduce MESH, a modularized unified retrieval architecture that decouples heterogeneous user, item, and context signals prior to shared interaction, enabling a single retrieval backbone to better preserve sparse-content learning signals while retaining efficiency and maintainability. 
    \item \textbf{Industrial validation at Pinterest scale}: We validate MESH on billion-scale Related Pins platform, demonstrating improved scaling behavior on fresh content offline and meaningful gains on fresh and long-tail segments in online A/B tests, alongside a 2.87$\times$ serving efficiency improvement.
\end{itemize}

By shifting from brute-force flat scaling to structured semantic scaling, we offer a practical blueprint for building unified retrieval backbones that scale more equitably across heterogeneous content.

\section{Methodology}
\subsection{The Scaling Bias of Heterogeneity}\label{method:bias}

The fundamental challenge in scaling discriminative recommendation models lies in maintaining signal fidelity across heterogeneous data distributions. In industrial platforms, the feature space $\mathcal{X}$ is composed of highly skewed partitions: $\mathcal{X} = \{ \mathcal{X}_{\mathrm{evergreen}} \cup \mathcal{X}_{\mathrm{fresh}} \cup \mathcal{X}_{\mathrm{tail}} \}$. In a traditional flat scaling framework, the model processes a "flat" set of features $E = [e_1, e_2, \dots, e_n]$ through a complex interaction backbone $H = \Phi(E; \Theta)$, where $\Phi$ represents high-order crossing layers (e.g., cross-networks, transformers, or hierarchical evolutions). The gradient of the loss $\mathcal{L}$ with respect to a specific item embedding $e_{\mathrm{item}}$ is governed by the chain rule:$$\nabla_{e_{\mathrm{item}}} \mathcal{L} = \frac{\partial \mathcal{L}}{\partial H} \cdot \frac{\partial \Phi(E; \Theta)}{\partial e_{\mathrm{item}}}$$

In architectures with explicit feature crossing (e.g., DCN\cite{wang2021dcn}, DHEN\cite{zhang2022dhen}), this coupling across the entire feature set ensures that global optimization is predominantly driven by high-frequency evergreen content. Consequently, the gradient signals for fresh entities are effectively masked by the head's statistical properties. As model capacity $N$ increases, this structural bias leads to interaction saturation for evergreen entities and signal suppression for the tail, formalizing the scaling disparity:$$\mathbb{E}_{\mathcal{X}_{\mathrm{fresh}}} [ \| \nabla_{e_{\mathrm{fresh}}} \mathcal{L} \| ] \ll \mathbb{E}_{\mathcal{X}_{\mathrm{evergreen}}} [ \| \nabla_{e_{\mathrm{evergreen}}} \mathcal{L} \| ]$$
This disparity creates \textit{the scaling bias of heterogeneity}: the backbone learns to progressively ignore sparse signals in favor of minimizing the global objective via dominant head-features, leading to the stagnant scaling exponents observed in our empirical study (Section \ref{results:fresh_scaling}). To mitigate this, we propose MESH framework (as shown in Figure \ref{fig:motivation}c), which acts as a structural inductive bias to shield fresh signals from global gradient interference. The structural design of MESH aims to ensure scaling fairness, where the capacity gains for fresh items are preserved relative to evergreen content.

%In a heterogeneous corpus, we identify the Gradient Dominance effect: because evergreen items $e_{\mathrm{head}}$ possess significantly higher co-occurrence frequencies and engagement densities, the attention scores $\text{Attn}(e_{\mathrm{head}}, e_j)$ exhibit much larger magnitudes and smaller variances compared to fresh items $e_{\mathrm{fresh}}$. Consequently, during the backpropagation of a massive Transformer, the update direction is effectively "hijacked" by the head content:$$\| \nabla_{e_{\mathrm{head}}} \mathcal{L} \| \gg \| \nabla_{e_{\mathrm{fresh}}} \mathcal{L} \|$$ 
% mitigated via residual connection or loss regularization etc.

\subsection{Structural Decoupling via Modularized Sub-Towers}
To counter scaling bias, MESH employs a structural decoupling strategy. The core philosophy is to isolate and enrich heterogeneous signals within independent, domain-specific sub-towers before they enter the global interaction space.
We partition the feature set $\mathcal{F}$ into $K$ semantic groups $\mathcal{G} = \{G_u, G_i, G_c\}$, representing user, item, and context domains.

\subsubsection{Group-wise Manifold Initialization}
Unlike flat architectures, we apply a group-wise inductive bias before any cross-domain interaction. Each group $G_k$ is first transformed into a dense manifold:$$H_k^0 = \text{LayerNorm}\left( \text{Concat}(\{e_j \mid f_j \in G_k\}) \right)$$
The initial LayerNorm within each group ensures that the internal variance of the Item domain is preserved independently of the user or context domains, preventing the evergreen frequency bias from polluting the initial latent representation of fresh items.

\subsubsection{Intra-tower Signal Amplification}

For sparse entities with limited historical interactions, raw embeddings are often insufficient to survive deep non-linear transformations, leading to semantic collapse. To preserve signal fidelity, we apply a domain-specific encoder $\psi_k$ (acting as a signal amplifier) to expand the raw group embeddings into a high-dimensional semantic super token $S_k$:
$$S_k = \psi_k(H_k^0; \Theta_k)$$

While MESH is agnostic to the specific choice of encoder, we instantiate $\psi_k$ using DHEN\cite{zhang2022dhen} due to their superior stability and expressive power in processing sparse categorical features. In this implementation, each sub-tower utilizes a recursive residual structure with hierarchical normalization. Each layer $l \in [1, L]$ performs a projection that repeatedly incorporates the original semantic signal to prevent information decay:
$$H_k^l = \text{Norm}\left( H_k^{l-1} + \Phi(H_k^{l-1}, H_k^{l-1}; \theta_k^l) \right)$$
where the $\Phi$ operator models explicit feature conjunctions including MaskNet, Transformer, DCNv2 and MLP.

This recursive mechanism functions as a signal amplifier: by repeatedly projecting the original semantic manifold $H_k^0$ back into the latent state, we ensure that intrinsic item features are fully expanded before they are exposed to the ``noise'' of global engagement biases in the crossing backbone.
This prevents the semantic collapse that often occurs in deep flat MLPs when dealing with sparse long-tail features and provides a protected gradient path for sparse entities, contributing to improved scaling fairness observed in our results (Section \ref{results:fresh_scaling}).

\subsection{Inter-group Global Crossing and Debias}
While the modularized sub-towers preserve the fidelity of domain-specific signals, the final interaction stage must remain resilient to the inherent engagement biases (e.g. device, time-of-day, user intent) prevalent in production environments. To prevent extrinsic context noise from polluting the intrinsic user-item matching, we propose a Residual Gated Bias Correction (RGBC) mechanism.

\subsubsection{Gated Bias Correction.}
To isolate the fundamental compatibility between the user and item, we first extract a clean latent affinity, independent of environmental factors. This is achieved via an element-wise Hadamard product of the user and item super tokens: $$\mathbf{X}_{ui} = T_u \odot T_i$$This term represents the intrinsic relevance manifold that MESH seeks to protect throughout the scaling process.
The context super token $T_c$ encapsulates environmental metadata that often correlates with engagement but not necessarily with content utility. We utilize $T_c$ to generate a confidence gate $\mathbf{g} \in (0, 1)$, which dynamically modulates the intensity of the semantic signal relative to the observed contextual bias: $$\mathbf{g} = \sigma \left( \text{MLP}_{\mathrm{gate}}(T_c) \right)$$where $\sigma$ denotes the Sigmoid activation function. The final input to the global crossing backbone, $\mathbf{H}_0$, is constructed by modulating the semantic core with this gate and incorporating a residual context bias, followed by a Layer Normalization step: $$\mathbf{H}_0 = \text{LayerNorm}( (\mathbf{g} \odot \mathbf{X}_{ui}) + T_c )$$ This design ensures that in high-bias scenarios, the context $T_c$ can absorb the engagement variance, allowing the user and item towers to focus purely on content utility.

\subsection{System Implementation \& Serving Efficiency}
To deploy MESH in a latency-critical production serving environment, we design a set of system-level optimizations that transform its structural modularity into inference efficiency gains.

\textit{Inter-Op Parallelization via TorchScript}. The modular design of MESH introduces structurally independent towers whose inference shares no data dependencies. In a naive PyTorch eager execution, these towers are executed sequentially due to Python-level dispatch and implicit synchronization, leading to under-utilization of GPU compute resources--particularly under small-batch serving workloads where kernel launch and scheduling overhead dominate. To address this inefficiency, we leverage TorchScript (Just-In-Time compilation) to enable inter-op parallelization across towers, allowing sub-towers to inference concurrently (Figure \ref{fig:motivation}c). This transformation enabled improved kernel overlap and reduced idle GPU cycles. Formally, let $T_k$ denote the inference latency of the $k$-th tower. Sequential execution incurs a backbone latency of $\sum_{k=1}^K T_k$, whereas our parallel execution reduces the critical path to $\max_{k}(T_k) + \epsilon$, where $\epsilon$ accounts for synchronization and scheduling overhead. In practice, this substantially improves effective GPU utilization, resulting in a near $K\times$ speedup in the encoding phase under serving conditions.

\textit{Threading Strategy and Resource Contention}. During deployment, we found that the host-side threading strategy for inter-op parallelism plays a critical role in tail latency (P99), especially under high-traffic conditions. Although assigning one dedicated thread per sub-tower appears intuitive, this approach led to frequent latency spikes in practice due to a cascade of context-switching overhead and resource contention. Utilizing a larger thread pool managed by the system's available compute resources effectively mitigated latency spikes. This strategy ensured a consistent flow of execution across the sub-towers, leading to more stable tail latency without sacrificing throughput.

% \subsection{Multi-Tower Contrastive Supervision}

% The modularity of MESH allows us to introduce auxiliary supervision signals that are impossible in flat architectures. We propose a Tower-level Contrastive Loss ($\mathcal{L}_{cl}$) to enhance semantic relevance:$$\mathcal{L}_{cl} = -\log \frac{\exp(\text{sim}(T_u, T_i) / \tau)}{\sum_{j \in \mathcal{B}} \exp(\text{sim}(T_u, T_{i,j}) / \tau)}$$where $T_u$ and $T_i$ are the User and Item Super Tokens, and $\mathcal{B}$ is the mini-batch. By aligning the tower embeddings in the latent space, we provide an additional "relevance gradient" that specifically benefits Fresh Items, as it forces the Item Tower to learn robust representations from content features even in the absence of click-through labels.

\section{Experiments}
\subsection{Experimental Setup}
\subsubsection{Model architecture and hyperparameters}
Our retrieval system follows a two-tower architecture, decoupling query and candidate representations to enable efficient approximate nearest neighbor (ANN) search. Architectural modifications are limited to the request-level (user-tower) representations. The flat architecture baseline concatenate all feature embeddings into a single vector prior to global DHEN interaction, while MESH introduces a modularized encoder that partitions heterogeneous signals into independent semantic towers followed by gated bias correction. Both architectures follow a consistent three-stage execution pipeline: (1) \textit{Feature Aggregation}: Sparse categorical features (e.g., semantic IDs, user/item IDs) are mapped to dense embeddings via learned lookup tables, while continuous features (e.g., engagement statistics) are concatenated. (2) \textit{Global Interaction}: A DHEN layer to capture high-order non-linearities. (3) \textit{Projection}: A series of fully connected layers with ReLU activation and Layer Normalization transform the interaction output into final L2-normalized embeddings $\mathbf{u}, \mathbf{v} \in \mathbb{R}^{64}$. To mitigate popularity bias during training, we incorporate in-batch negatives with logit correction\cite{yi2019sampling} based on item frequency.

% \red{"Flat-DHEN: A non-modularized version of the DHEN architecture, where user, item, and context features are flattened into a single input layer, bypassing the structural decoupling of MESH. This serves as our primary baseline to isolate the impact of modularization."}
% (i.e., monolithic backbones where all feature embeddings are concatenated into a single vector prior to global interaction)

\begin{table}[b]
    \caption{Corpus selection and pruning logic across content segments}
    \centering
    \renewcommand{\arraystretch}{1.3}
    \small
    \begin{tabular}{lp{55mm}}
        \toprule
        \textbf{Corpus Type} & \textbf{Conceptual Selection / Pruning Logic} \\
        \midrule
        \textbf{Fresh Cold-start} & \textbf{Select:} Items within a brief ingestion window $\Delta T_{\mathrm{new}}$ with minimal exposure $I < \tau_{\mathrm{exp}}$. \\
                                 & \textbf{Prune:} Items exceeding a maximal latency window without meeting engagement minimums. \\
        \midrule
        \textbf{Fresh Warm-start} & \textbf{Select:} Content showing initial traction ($I \ge \tau_{\mathrm{exp}}$) and meeting a quality prior based on relative conversion rates. \\
                                 & \textbf{Prune:} High-impression items that fail to sustain the engagement density $\rho_{\mathrm{min}}$. \\
        \midrule
        \textbf{Shopping Long-tail} & \textbf{Select:} Items identified via semantic shopping intent indicators $\mathbb{I}_{\mathrm{shop}}$. \\
        \midrule
        \textbf{Evergreen}       & \textbf{Select:} Highly-engaging content identified by a cumulative engagement score with a temporal decay factor: $S_{\mathrm{evg}} = \sum v_i \cdot e^{-\lambda(T - t_i)}$. \\
        \bottomrule
    \end{tabular}
    \label{tab:corpus}
\end{table}

\subsubsection{Training and Evaluation Data.} We collect 15 days of user engagement logs from Pinterest's Related Pins platform, comprising query-candidate pairs where users engaged with recommended pins. Data is featurized by joining with inference logs to obtain item features, user context, and engagement labels. We end up with around 1B records for training after various sampling strategies. 
To ensure an unbiased assessment, the offline evaluation set is constructed from a separate 3-day window following the training period, utilizing a 5\% off-policy uniform random traffic slice. This yields 450 million test examples, providing a statistically robust foundation for offline benchmarking. To evaluate scaling fairness, we stratify the test set into evergreen and fresh corpora (age $\le$ 28 days), allowing for a granular analysis of cold-start performance across different content life-cycles. Specifically, fresh-content metrics (e.g., Recall@$K$) are computed exclusively against the fresh candidate subspace, preventing evergreen head entities from diluting the performance signal.

While our evaluation relies on proprietary Pinterest data, we argue that this is the most suitable setting for studying the Scaling Bias of Heterogeneity. Public benchmarks (e.g. MovieLens, Amazon Reviews) lack two properties essential to our analysis: (1) a natural stratification of content into heterogeneous tiers with vastly different interaction densities (evergreen vs. fresh vs. long-tail), and (2) sufficient scale to fit reliable power-law curves across multiple FLOPs configurations — these datasets saturate at low model capacity, making scaling law analysis statistically fragile. In contrast, our offline evaluation is constructed from a 5\% off-policy uniform random traffic slice, which eliminates selection bias introduced by the production ranking policy and provides an unbiased estimate of retrieval quality. Combined with 3-week billion-scale online A/B tests, we believe this constitutes a rigorous and appropriate validation setting for this class of problem.

\subsubsection{Offline Metrics.}
We employ recall@$K$ as our primary offline metric, defined as the fraction of ground-truth relevant items retrieved within the top-$K$ candidates: $$\text{Recall@K} = \frac{|S_K \cap R|}{|R|}$$
where $S_K$ denotes the set of top-$K$ items retrieved by the model, and $R$ denotes the set of ground-truth relevant items for the given query pin. Among various interaction types, we choose repin (user saves pin to their boards) as the primary positive signal for training, as it represents the strongest indicator of user interest for organic content.
We focus on Recall@10 as the primary offline metric due to its strong correlation with online engagement. 
To ensure statistical stability during scaling law characterization, we prioritize the analysis of two distinct content archetypes: evergreen and fresh. We intentionally exclude deep-tail shopping segments from the offline scaling analysis, as their extreme signal sparsity often leads to high-variance metrics that hinder reliable power-law fitting.
We adopt the fresh segment as a high-entropy proxy for the broader tail, providing the signal density necessary for stable scaling analysis. Online evaluations (Section \ref{results:online_overall}) confirm these gains successfully generalize to the commercial shopping tail.

\begin{table}[b]
    \caption{Systematic evaluation of MESH Scaling efficiency across architectural dimensions.}
    \centering
    \renewcommand{\arraystretch}{1.1}
    \small
    \begin{tabular}{
        >{\raggedright\arraybackslash}p{25mm}   % Configuration
        >{\centering\arraybackslash}p{10mm}      % FLOPs
        >{\centering\arraybackslash}p{10mm}     % AUC
        >{\centering\arraybackslash}p{15mm}     % LogLoss
        >{\centering\arraybackslash}p{10mm}     % Delta AUC
    }
        \toprule
        \textbf{Configuration} & \textbf{TFLOPs} & \textbf{Params} & \textbf{Recall@100} & \textbf{$\Delta$Recall} \\
        \midrule

        \multicolumn{5}{p{65mm}}{\textit{Varying per field feature dimension $d$}} \\
        d = 16         & 2.8 & 17.8M & 0.835 & 0.00\% \\
        d = 32         & 3.2 & 25.7M & 0.870 & +4.19\% \\
        d = 64         & 3.4 & 48.0M & 0.886 & +6.11\% \\
        d = 128         & 6.6 & 118.2M & 0.888 & +6.35\% \\
        \midrule

        \multicolumn{5}{p{65mm}}{\textit{Varying output per field dimension $L$}} \\
        L = [16, 16]         & 3.2 & 25.7M & 0.870 & 0.00\% \\
        L = [32, 32]         & 4.2 & 43.8M & 0.873 & +0.34\% \\
        L = [64, 64]         & 6.0 & 105.0M & 0.882 & +1.38\% \\
        L = [128, 128]         & 10.8 & 328.2M & 0.893 & +2.64\% \\
        \midrule

        \multicolumn{5}{p{65mm}}{\textit{Widening interaction modules per DHEN layer $C$}} \\
        C = [16]         & 2.0 & 7.3M & 0.866 & 0.00\% \\
        C = [16, 16]     & 3.2 & 25.7M & 0.870 & +0.46\% \\
        C = [16, 16, 16] & 3.3 & 29.3M & 0.882 & +1.85\% \\
        C = [16, 16, 16, 16]  & 3.4 & 34.7M & 0.883 & +1.96\% \\
        \bottomrule
    \end{tabular}
    \label{tab:dhen_scaling}
\end{table}

\subsubsection{Online A/B tests.} We deploy MESH in Pinterest's production Related Pins system and conduct large-scale A/B experiments to evaluate real-world impact. The experiments run for 3 weeks with traffic split between control (baseline flat model in production) and treatment groups using consistent user-level randomization. The control group uses the production retrained model of two-tower retrieval model with same dates of training data. We measure related-pins overall repin rate (our primary organic content quality metric) and segmented content engagement to evaluate cold-start improvements, while monitoring guardrail metrics (P99 latency, infrastructure cost) to ensure production stability.

\subsubsection{Heterogeneous corpus tiering and selection} To evaluate the generalizability of MESH across varying signal densities, we partition the production content ecosystem into a multi-tier hierarchy based on temporal dynamics and interaction sparsity (Table \ref{tab:corpus}). For the fresh segment, we follow a graduation-based trajectory\cite{wang2023fresh} where \textit{Fresh Cold-start} items (newly published, near-zero signals) transition into \textit{Fresh Warm-start} once they surpass specific engagement thresholds. In parallel, we isolate shopping items which represent a unique long-tail semantic distribution characterized by high commercial intent but sparse interaction density. Finally, the evergreen corpus—comprising the stable platform "head" with high engagement rate—serves as the benchmark for system stability and no-regression verification during scaling. While the underlying structural logic remains constant, the specific selection thresholds (e.g., $\Delta T_{\mathrm{new}}$, $\tau_{\mathrm{exp}}$) are adaptively tuned to reflect shifting platform traffic patterns.

% EBIS score -- a learned, content-based engagement score used as a proxy for how engaging a pin is.
% PLP -- organic, non-product pins that have shoppable product associations 

\subsection{Discriminative Scaling Law}

\subsubsection{General Scaling efficiency.}\label{results:general_scaling}
We first evaluate the fundamental scaling properties of MESH framework by systematically varying its structural hyperparameters: input feature/token dimension ($d$), output embedding dimension of DHEN layers ($L$), and the number of interaction modules per DHEN layer ($C$). We model the relationship between computational budget and performance using the power-law form: $$y = \alpha x^\beta + \gamma$$
where $y$ represents the performance metric and $x$ denotes the scaling factor (FLOPs or parameter count), and $\beta$ is the scaling exponent. 
As summarized in Table \ref{tab:dhen_scaling}, MESH demonstrates consistent performance gains as model capacity increases. Specifically, expanding the model's width ($d, L$) and depth ($C$) leads to a robust upward trend in retrieval precision. While traditional models often hit a performance ceiling due to feature saturation, our structured modular design shows a more consistent improvement trend. However, as $x$ increases, the marginal gain in Recall@10 begins to slow, reflecting the transition towards the diminishing returns regime. 
%However, this global saturation is largely driven by the informational ceiling of the evergreen majority and does not imply a lack of growth in underserved segments. 
%To reveal the continued growth within underserved segments, we next stratify these results to evaluate \textit{scaling fairness}.

\begin{figure}[b]
  \centering
  \includegraphics[width=\linewidth]{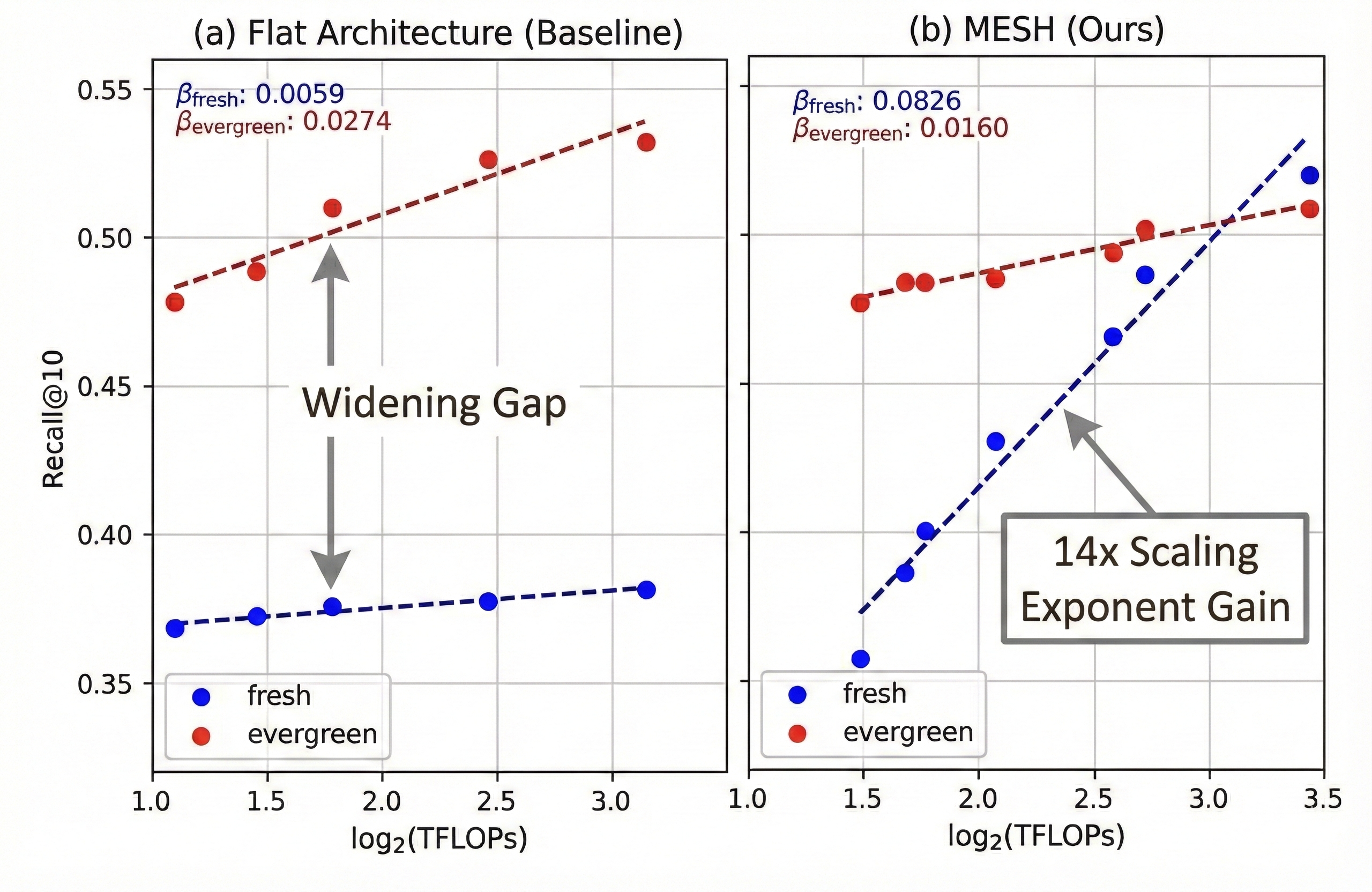}
    \caption{Breaking the Scaling Bias. (a) "Flat" Architecture (Baseline): As compute budget increases, the scaling trajectory is dominated by evergreen content, leading to a stagnant $\beta_{\mathrm{fresh}} = 0.0059$ and a widening representation gap. (b) MESH framework: By decoupling the gradient paths, MESH recovers the latent scaling potential of the fresh segment, achieving a 14$\times$ improvement in scaling exponent ($\beta_{\mathrm{fresh}} = 0.0826$) and effectively closing the performance gap between head and tail entities. Each data point represents the mean of five independent evaluations to ensure statistical robustness.}
    \label{fig:segment_scaling}
\end{figure}

\subsubsection{Evaluating Scaling Fairness}\label{results:fresh_scaling} A central premise of our work is that model capacity should translate into performance gains across the entire content ecosystem, ranging from evergreen to fresh corpus. However, our empirical analysis reveals that traditional ``flat-architectures'' suffer from \textit{Scaling Divergence} when faced with heterogeneous data distributions. As illustrated in Figure \ref{fig:segment_scaling}, the scaling trajectory of the "flat architecture" baseline is dominated by a pervasive ``head-bias''. The scaling exponent $\beta$ for the evergreen corpus—defined by the power-law relationship $\mathrm{Metric} \propto N^{\beta}$—remains relatively robust ($\beta_{\mathrm{evergreen}} = 0.0274$). In contrast, the $\beta$ for the fresh segment is nearly stagnant ($\beta_{\mathrm{fresh}} = 0.0059$). This disparity confirms the gradient dominance hypothesis formulated in Section \ref{method:bias}: in a globally coupled architecture, additional compute is disproportionately consumed by high-frequency entities, while the sparse signals of fresh content remain effectively masked within the interaction layers. Consequently, in traditional models, heterogeneity scales inefficiently; the representation gap between the dominant (head) and sparse (tail) corpora widens as model size increases, reinforcing a structural rich-get-richer feedback loop.

%This phenomenon occurs because traditional models optimize a global loss function where the gradient is dominated by high-frequency evergreen entities. Consequently, for traditional models, heterogeneity does not scale; the gap between the "rich" (head) and "poor" (tail) corpora only widens as the model size increases.

In contrast, MESH demonstrates improved scaling behavior for tail segments. By decoupling the feature space into modularized sub-towers and enforcing structural induction via the RGBC mechanism, MESH reduces gradient dominance on the fresh segment, leading to a substantial improvement in its scaling efficiency ($\beta_{\mathrm{fresh}} = 0.0826$, a 14$\times$ improvement over baseline), as seen in Figure \ref{fig:segment_scaling}b. Remarkably, under the MESH framework, the fresh segment now exhibits an even steeper slope than the evergreen segment ($\beta_{\mathrm{evergreen}} = 0.0160$), suggesting that while evergreen representations are approaching a saturation point, the fresh segment possesses considerably higher scaling elasticity. The observed moderation in $\beta_{\mathrm{evergreen}}$ does not reflect a sacrifice in performance, but rather a more efficient reallocation of representational capacity. MESH effectively reallocates the model's capacity from redundant head-fitting to the high-growth potential of the content tail, achieving more efficient and diverse ecosystem scaling. By recovering the latent scaling potential of the tail, MESH helps ensure that increases in FLOPs contribute more equitably to ecosystem diversity, as validated by our online A/B tests (Section \ref{results:ab_test}).

To directly validate the ``protected gradient path'' mechanism, we measure per-group gradient statistics during training. In the flat baseline, evergreen L2 gradient norms are 12.4$\times$ larger than fresh ones, directly confirming signal masking. Under MESH, this ratio improves by 62\% (to 4.7$\times$), demonstrating that modular isolation actively preserves learning signals for sparse entities. Furthermore, using Integrated Gradients, content feature attribution scores increase by 28\% in MESH relative to the flat baseline, confirming a structural shift from popularity bias toward intrinsic semantic utility. To quantify the statistical robustness of these estimates, we bootstrap the power-law fit with 1,000 resamples. The 95\% confidence interval for $\beta_{\text{fresh}}$ under MESH is $[0.0772, 0.0874]$, with the lower bound remaining ${\sim}13{\times}$ the flat baseline ($\beta_{\text{fresh}} = 0.0059$), confirming that the 14$\times$ scaling exponent gain is statistically robust and not an artifact of sparse data points.

\begin{table}[h]
    \caption{Ablation study on key components of MESH.}
    \centering
    \renewcommand{\arraystretch}{1.2}
    \small
    \begin{tabular}{
        >{\raggedright\arraybackslash}p{42mm}
        >{\centering\arraybackslash}p{14mm}     
        >{\centering\arraybackslash}p{8mm}     
        >{\centering\arraybackslash}p{8mm}        
    }
        \toprule
        \textbf{Configuration} & Overall Recall@10 & \textbf{$\beta_\mathrm{fresh}$}  & $\Delta \beta_\mathrm{fresh}$ \\
        \midrule
        \textbf{MESH framework}      & \textbf{0.521} & \textbf{0.0826} & - \\
        \quad -- w/o Semantic Amplifiers    & 0.482 & 0.0071 & -91.4\% \\
        \quad -- w/o Modular Crossing Backbone & 0.495 & 0.0528 & -36.1\% \\
        \quad -- w/o RGBC mechanism         & 0.509 & 0.0645 & -21.9\% \\
        \bottomrule
    \end{tabular}
    \label{tab:offline_ablation}
\end{table}

\subsubsection{Ablation study.} 

To isolate the contributions of MESH’s architectural components to its heterogeneous scaling, we conduct a series of ablation experiments summarized in Table \ref{tab:offline_ablation}.
The most significant performance degradation occurs when the semantic amplifiers are replaced with traditional flat embedding layers; in this configuration, the scaling exponent $\beta_{\mathrm{fresh}}$ collapses from 0.0826 back to 0.0071, confirming that modular inductive bias at the input level is the primary driver of recovery by preventing "semantic washout". Next, ablating the modular crossing backbone by replacing DHEN-based layers with a MLP of equivalent parameter size leads to a substantial degradation in $\beta_{\mathrm{fresh}}$. This suggests that while structured inputs provide the opportunity to scale, the hierarchical residual connections in the backbone are essential for gradient protection. Finally, evaluation of the RGBC mechanism reveals that replacing the gating layer with a standard addition operation leads to a noticeable drop for the fresh segments. This validates the necessity of a signal purifier to modulate contextual noise and ensure scaling fairness, as without RGBC, increased model capacity is partially wasted on fitting extrinsic engagement patterns rather than intrinsic compatibility. 

We also evaluate the sensitivity of MESH to the number of sub-towers $K$. The choice of $K=3$ reflects the three core semantic domains in retrieval: user, item, and context. Testing $K=4$ by further partitioning the query pin sub-tower into engagement and content sub-towers yields marginal gains of $+0.3\%$ fresh impressions and $+0.5\%$ repins over $K=3$. This confirms that performance scales with the semantic coherence of the partitioning rather than cardinality alone. We chose $K=3$ as the production sweet spot, as the marginal gains of $K=4$ did not justify the added serving complexity under high-QPS constraints. Similarly, MESH is encoder-agnostic by design: preliminary experiments with MLP and DCNv2 encoders demonstrate consistent scaling fairness gains, though with smaller magnitudes than DHEN, confirming that modular domain isolation is the primary driver of scaling fairness independent of encoder choice.

%Table \ref{tab:offline_ablation} presents an ablation study on the key components in MESH. We first examine the impact of the semantic grouping on super tokens. Compared to the base model without merging, integrating semantic tokens reduces FLOPs from 3.73$\times$10$^9$ to 3.03$\times$10$^9$ while improving recall@10 by 1.58\% and decreasing loss by 3.48\%. Further incorporating DHEN as intra-group amplifier brings additional gains, achieving the best overall LogLoss of 0.47052 and a 1.63\% recall@10 improvement.

% We verify the scalability of MESH by measuring the relationship between the test cross-entropy loss $L$ and the total parameter count $N$.
% Hypothesis: $L(N) \propto N^{-b}$. We expect MESH to yield a higher power-law exponent $b$ for the "Fresh" subset of the corpus compared to flat Transformer baselines.Setup: We scale the hidden dimensions of the DHEN towers from $256$ to $4096$ and the number of layers from $2$ to $12$.

\subsection{Online A/B Tests Results}\label{results:online_overall}
\subsubsection{Segment-specific Lifts and Ecosystem Impact}\label{results:ab_test}
To evaluate the real-world impact of MESH, we conducted large-scale online A/B tests on a production Related-Pins platform, focusing on its ability to handle a highly heterogeneous content ecosystem. The experimental results reveal a non-cannibalizing scaling effect -- MESH achieves significant gains in underserved segments without regressing on global performance.

\begin{figure}[t]
  \centering
  \includegraphics[width=\linewidth]{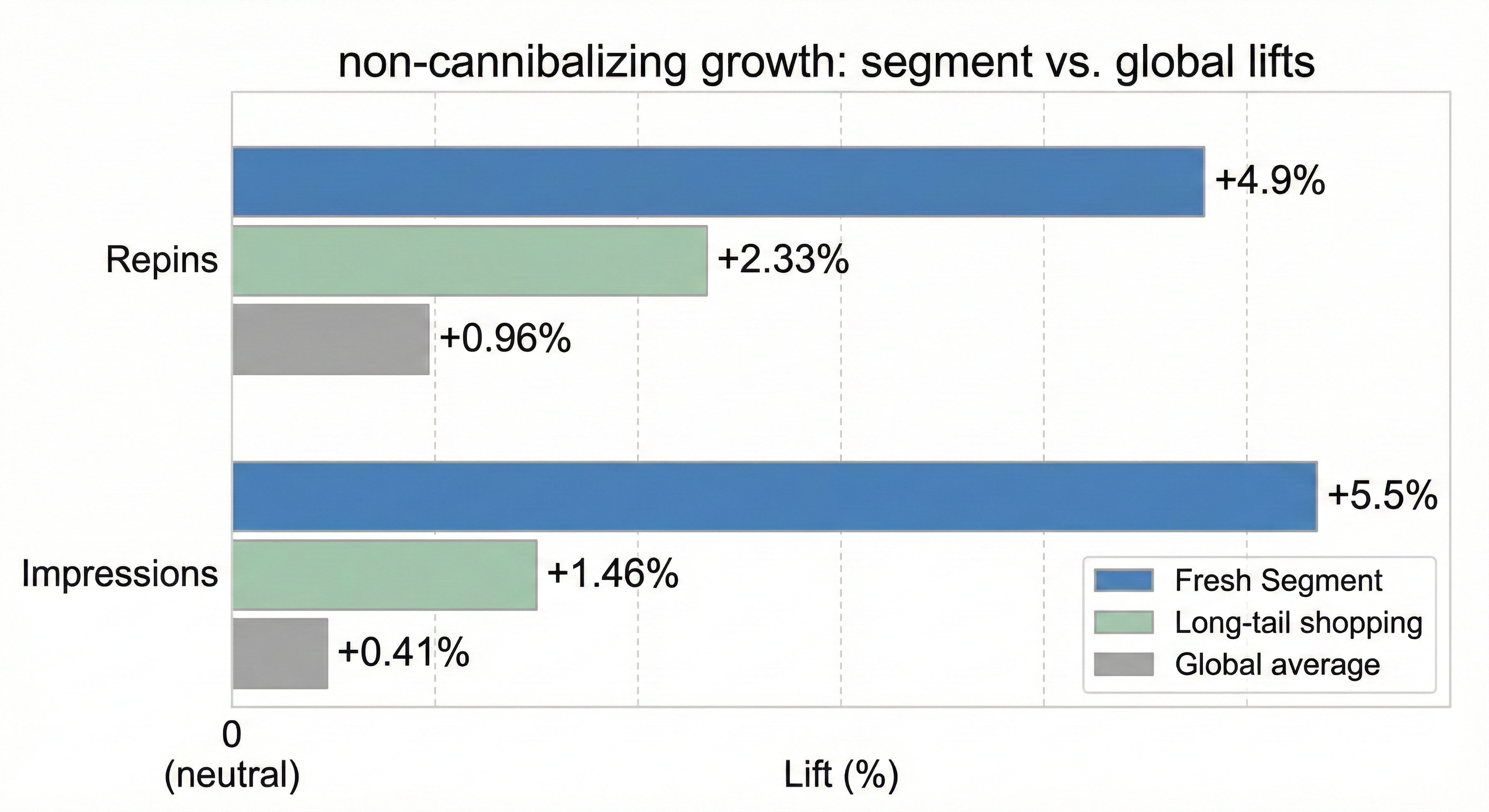}
    \caption{Non-cannibalizing growth through MESH framework. Online A/B test results demonstrate significant engagement (Impressions and Repins) surges in fresh and long-tail shopping segments.}
    \label{fig:fresh_metric}
\end{figure}

For the fresh Segment, MESH achieved substantial gains with a +5.5\% lift in impressions and a +4.9\% increase in repins as shown in Figure \ref{fig:fresh_metric}. As detailed in Table \ref{tab:age_decomposed}, these gains are primarily driven by the 7--28 day warm-start window, where the model successfully recovers latent signals. Furthermore, long-tail shopping pins—a critical segment for e-commerce discovery—saw significant improvements, with impressions rising by +1.46\% and repins by +2.33\%. %These results prove that MESH successfully extracts high-utility representations where traditional monolithic backbones typically struggle due to gradient dominance from popular content.
Crucially, these targeted gains were achieved while maintaining overall system stability, as evidenced by the neutral-to-positive lifts in total impressions (+0.41\%) and total repins (+0.96\%). In the context of billion-scale recommendation engines, an overall neutral result paired with segment-specific gains reflects a more equitable distribution of traffic. It demonstrates that MESH does not merely shift engagement from one pool to another via cannibalization. Instead, it improves retrieval coverage by surfacing content that was previously underserved by the legacy backbone. By breaking the rich-get-richer loop identified in Section \ref{results:fresh_scaling}, MESH supports a more diverse content lifecycle.

To further validate that these gains reflect a structural advantage over external scaling-aware architectures, we conduct online A/B tests comparing MESH against RankMixer~\cite{zhu2025rankmixer} — a SOTA scaling-aware flat architecture — at two scales (MESH as baseline): RankMixer-small (28M) yields $-$1.42\% fresh repin and $-$1.18\% fresh impression; RankMixer-large (135M) yields $-$2.99\% and $-$1.77\% respectively. Crucially, scaling RankMixer widens the performance gap relative to MESH rather than closing it, confirming that the Scaling Bias is a structural problem that flat architectures cannot resolve through brute-force capacity expansion alone.

\begin{table}[ht]
    \caption{Age-band decomposed freshness impact from online A/B tests. \textbf{Statistically significant} results are bolded.}
    \centering
    \renewcommand{\arraystretch}{1.3}
    \small
    \begin{tabular}{lcccc}
        \toprule
        \textbf{Pin Age} & \textbf{0--7 days} & \textbf{7--14 days} & \textbf{14--28 days} & \textbf{28--90 days}\\
        \midrule
        Impressions & +0.11\%  & \textbf{+1.60\%} & \textbf{+2.00\%} & +0.55\% \\
        Engagement & +0.96\% & \textbf{+2.49\%} & \textbf{+2.77\%} & +0.56\% \\
        \bottomrule
    \end{tabular}
    \label{tab:age_decomposed}
\end{table}

\subsubsection{Ecosystem-wide Impact on Funnel Efficiency}
Beyond the immediate increase in content-specific metrics, we investigate whether the improved distribution of fresh and long-tail content translates into a healthier global ecosystem. To evaluate system-wide impact on the multi-stage recommendation funnel, we benchmark MESH's performance across heterogeneous corpora using two key industrial metrics: Survival Rate and CG Efficiency. These metrics are collected from logging of online experiment data. To quantify the health of the candidate generation (CG) stage, we define:
\begin{itemize}
    \item \textit{Survival Rate (SR)}: Measures the proportion of nominated candidates that successfully pass the final ranking stage.$$SR = \frac{|\mathcal{C}_{\mathrm{ranked}}|}{|\mathcal{C}_{\mathrm{total\_nominations}}|}$$
    \item \textit{CG Efficiency (CE$@$50)}: Measures the proportion of nominated candidates that reach the top-50 final ranked list, representing the "high-utility" retrieval rate.$$CE@50 = \frac{|\mathcal{C}_{\mathrm{ranked}} \cap \mathcal{P}_{50}|}{|\mathcal{C}_{\mathrm{total\_nominations}}|}$$
\end{itemize}
A higher SR indicates a better alignment between retrieval and ranking, while a higher CE$@$50 signifies the model's ability to surface high-utility items that provide immediate value.

\begin{figure}[b]
  \centering
  \includegraphics[width=\linewidth]{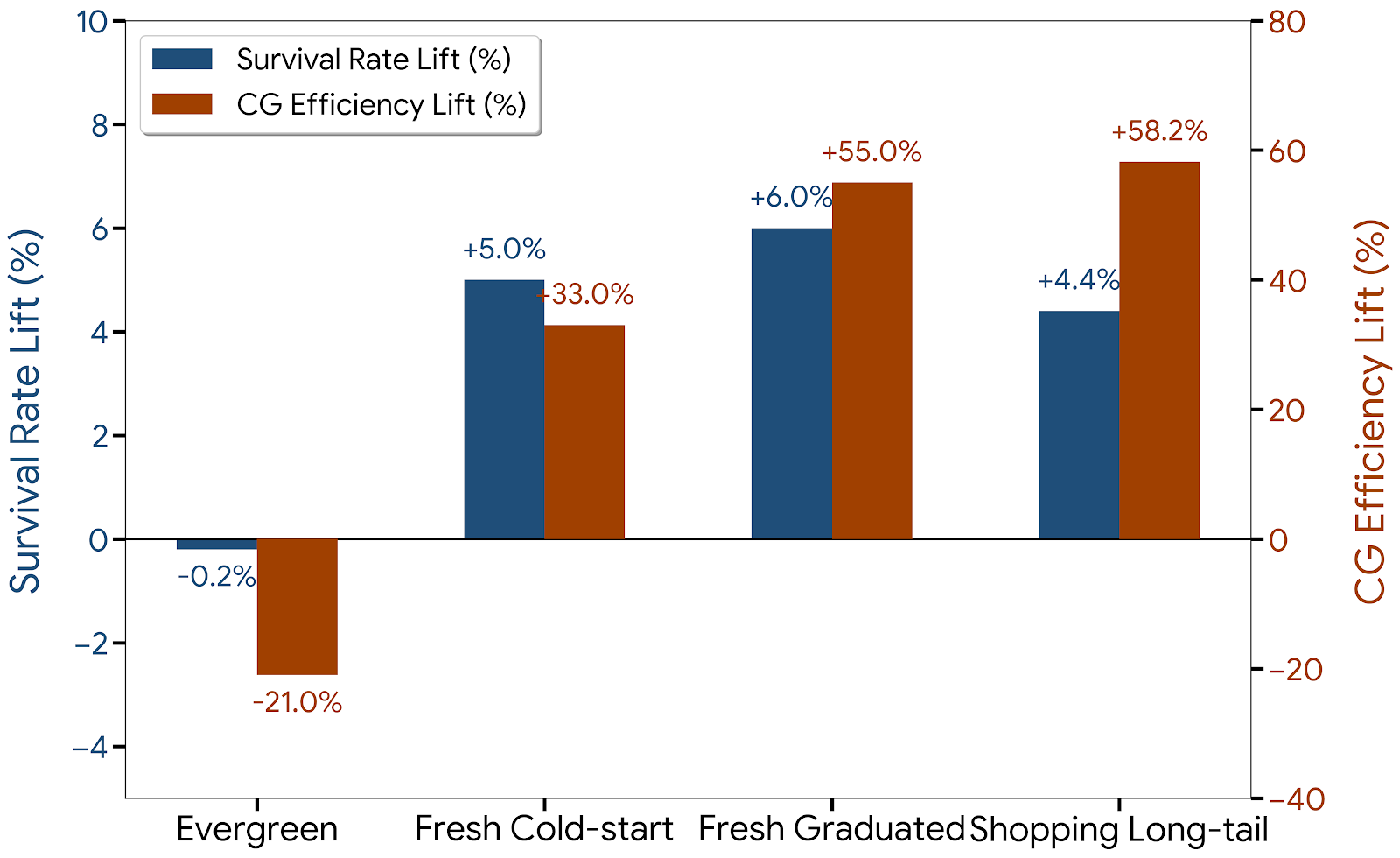}
    \caption{Funnel efficiency gains across heterogeneous corpora. The dual-axis visualization highlights the significant lift in fresh and long-tail segments while maintaining stability in the evergreen corpus.}
    \label{fig:cg_efficiency}
\end{figure}

As illustrated in Figure \ref{fig:cg_efficiency}, the MESH framework exhibits a substantial impact on the discovery ecosystem. We observe significant improvements in funnel penetration across underserved segments: fresh warm-start items achieved a +55.0\% lift in CE and a +6.0\% lift in SR, while shopping long-tail pins saw the most significant CE surge of +58.2\% alongside a +4.36\% SR lift. These results empirically demonstrate that MESH better preserves sparse signals that were previously suppressed. By providing a "protected gradient path" through modular sub-towers, MESH ensures that high-quality fresh and long-tail items can effectively compete in the final ranking stages, ultimately achieving better model scaling fairness.

Regarding the evergreen corpus, the transition resulted in a neutral SR (-0.21\%) and a 21.0\% decrease in CE. We interpret this not as a regression, but as a strategic architectural trade-off to combat evergreen over-fitting. In this high-frequency head segment, traditional architectures often over-optimize for hyper-popular items, leading to exposure redundancy. The reduction in CE for Evergreen, paired with a stable SR, suggests that MESH is diversifying the candidate pool by reallocating the system's discovery capacity toward high-potential fresh and shopping content. This redistribution prioritizes long-term ecosystem health over the redundant over-exposure of established head entities, suggesting a more equitable allocation of discovery capacity across content tiers.

% \subsubsection{Ablation Study: Impact of Contrastive Supervision}
% To evaluate the effectiveness of the Tower-level Contrastive Loss ($\mathcal{L}_{cl}$), we conducted an ablation study on the production dataset. As shown in Table 1, the inclusion of $\mathcal{L}_{cl}$ significantly improves metrics for the long-tail.

% Rationale: The gain in Fresh-AUC is notably higher than the global gain. This confirms our hypothesis: by aligning the User and Item Super Tokens through contrastive learning, the model forces the Item Tower to extract semantic relevance from raw content features. This provides a "warm-start" signal even for items with zero historical interactions, directly boosting discovery efficiency.

\subsubsection{User Retention Impact}
Online A/B testing indicates that the MESH framework yields concurrent improvements in user engagement and long-term retention metrics. As summarized in Table~\ref{tab:growth_metrics}, the model achieves a +0.51\% increase in Related Pins time spent, while simultaneously driving gains in Search discovery (+0.49\%) and Homefeed (+0.41\%). This concurrent growth across surfaces suggests that MESH's gains are not achieved at the expense of other surfaces. To evaluate the long-term value of these distribution shifts, we analyze deep engagement metrics as proxies for user retention. The framework achieves a +0.46\% lift in Sessions $>$ 5 mins, indicating a substantial increase in high-quality time spent. Combined with a +0.19\% increase in Total Successful Sessions (SSv2), these results suggest that MESH's improvements extend beyond immediate engagement metrics to longer-term user retention.

\begin{table}[ht]
    \caption{Impact on Growth Ecosystem Metrics. Statistically significant results are bolded.}
    \centering
    \renewcommand{\arraystretch}{1.3}
    \small
    \begin{tabular}{llcl}
        \toprule
        \textbf{Category} & \textbf{Metric} & \textbf{Online Impact} \\
        \midrule
        \multirow{2}{*}{\textbf{User Retention}} & Sessions $>$ 5 mins & \textbf{+0.46\%}  \\
                                                 & Successful Sessions (SSv2) & \textbf{+0.19\%} \\
        \midrule
        \multirow{3}{*}{\textbf{User Engagement}}  & Total time spent in App  & \textbf{+0.45\%} \\
                                                 & Time Spent (Related Pins) & \textbf{+0.55\%} \\
                                                 & Time Spent (Search) & \textbf{+0.49\%} \\
                                                 & Time Spent (Home Feed) & +0.41\% \\
        \bottomrule
    \end{tabular}
    \label{tab:growth_metrics}
\end{table}

% SSv2:
% Total Successful Sessions (SSv2): +0.19% (UCAN)
% Non Shopping Click Out Sessions, MECE (SSv2): +0.40% prop. (UCAN)

% GPS:
% Shop SRSD onlyer: +0.35% (UCAN) / +0.24% (all)

% Retention & Key Actions:
% Session > 5 mins:  +0.46% (UCAN)
% Related Pins: +0.51% (UCAN)
% Search: +0.49% (UCAN) / +0.30% (all)
% Homefeed: +0.41% (UCAN) / +0.03% prop. (all)

% Web Or API Requests: +0.28% (P6) / +0.36% (UCAN)
% Pin Impressions (1 pixel, 1 second): +0.34% (all) / +0.49% (UCAN)
% Ads Impressions: +0.39% (UCAN)
% Pin Searches: +0.15% (all) / +0.43% (P6) / +0.41% (UCAN)

\subsection{System Efficiency and Scalability}
To support billion-scale retrieval in high-QPS production environments, we evaluate the impact of inter-op parallelization and threading strategies on the inference performance of the MESH framework. The modular architecture of MESH facilitates high-concurrency execution. By utilizing TorchScript JIT to optimize the execution graph, we allow independent sub-towers to run asynchronously. As shown in Table~\ref{tab:latency}, a naive sequential implementation yields a P99 latency of 42.5~ms per forward pass. Notably, a sub-optimal parallelization strategy (e.g., $n=3$, matching sub-towers) induces severe resource contention, causing latency to spike to 187.4~ms. However, by aligning inter-op parallelism with the hardware capacity (48 threads), we achieve a near-optimal latency of 14.8~ms. This represents a 2.87$\times$ speedup over the sequential baseline, effectively decoupling total model capacity from linear latency accumulation. During production rollout, the MESH framework demonstrated stable CPU utilization with no significant increase in contention or wait times. This serves as empirical evidence that our modular design effectively utilizes hardware parallelism, ensuring that the system remains scalable even as model complexity increases.

\begin{table}[ht]
    \caption{Inference Performance across Execution Strategies.}
    \centering
    \renewcommand{\arraystretch}{1.2}
    \small
    \begin{tabular}{lccc}
        \toprule
        \textbf{Execution Strategy} & \textbf{Threads ($n$)} & \textbf{P99 Latency (ms)} & \textbf{Speedup} \\
        \midrule
        Sequential Baseline & 1 & 42.5 & 1.00$\times$ \\
        Sub-optimal Parallelism & 3 & 187.4 & 0.08$\times$ \\ 
        \textbf{MESH (Optimized)} & \textbf{48} & \textbf{14.8} & \textbf{2.87$\times$} \\
        \bottomrule
    \end{tabular}
    \label{tab:latency}
\end{table}

\section{Related Work}
\textbf{Discriminative Scaling Laws in Recommendation}. The success of LLMs has motivated a search for similar scaling laws in recommendation systems. Early explorations by DRLM\cite{zhou2018deep} and DIN\cite{naumov2019deep} demonstrated that increasing model depth and embedding dimensions leads to predictable gains in CTR prediction. More recently, works such as RankMixer\cite{zhu2025rankmixer} and OneTrans\cite{zhang2025onetrans} have expanded this paradigm by adopting a ``Flat Tokenization'' approach, treating heterogeneous features as a uniform sequence to be processed by massive Transformer backbones. While these models prove that ``more is better'' for global metrics, they frequently overlook the intra-corpus distributional shift. Our work argues for structural modularity as a prerequisite for equitable scaling across evergreen, fresh, and long-tail content.\\

\noindent \textbf{Model Debiasing and Unification for Cold Start}.
The industrial trend in recommendation is shifting from fragmented, task-specific models toward unified backbones\cite{chai2025longer,zhang2022dhen}. 
Historically, addressing engagement bias (e.g., popularity or position bias) and cold-start issues required maintaining a ``zoo'' of specialized models or auxiliary tasks\cite{guo2019pal, lee2019melu}. While two-tower retrieval architectures\cite{yi2019sampling} provide an inherent inductive bias by separating user and item representations, they often lack the expressive capacity of ranking models.
MESH bridges this gap by unifying specialized debiasing and universal scaling into a single cold-start-aware retrieval backbone.\\

\noindent \textbf{Feature Interaction and High-Order Operators}. 
The evolution of feature interaction has progressed from factorized parameters in sparse settings (FM \cite{rendle2010factorization}, FFM \cite{juan2016field}) to deep architectures like DCNv2 \cite{wang2021dcn} and Auto-Int \cite{song2019autoint} that capture explicit and self-attentive interactions. More recently, DHEN \cite{zhang2022dhen} achieved state-of-the-art performance by utilizing a recursive crossing mechanism for complex non-linearities. However, these operators are predominantly applied in a flat manner, leading to gradient dominance for sparse entities. We extend this paradigm by repurposing DHEN as a hierarchical inter- and intra-group encoder to isolate and amplify sparse signals, thereby ensuring more equitable representation across the corpus.\\

\section{Conclusion}
In this work, we characterized \textit{The Scaling Bias of Heterogeneity} as an important challenge to consolidating industrial-scale retrieval. While scaling laws in ranking models are typically observed within a curated subspace of high-quality candidates, retrieval systems must navigate the extreme entropy and sparse signal density inherent to a massive global, multi-tier corpus. Our findings suggest that balanced scaling across heterogeneous content tiers is important for unified retrieval systems. The proposed solution, MESH, represents a meaningful step toward consolidating fragmented model zoos toward a unified and scalable architecture. By establishing a robust, debiased foundation, MESH not only simplifies operations but also opens new directions for retrieval innovation. In terms of future work, the modular semantic-tower design provides the architectural flexibility to transcend traditional dot-product constraints, paving the way for sophisticated neural retrieval and complex interaction layers. Furthermore, by structurally decoupling semantic domains, MESH facilitates strategic hybrid inference that allows high-capacity item towers to be offloaded to offline batching or caching, supporting more efficient and sustainable large-scale unified retrieval. As we move toward an era of generative and multi-modal discovery, MESH's principles of modularity and scaling fairness offer a practical architectural framework for building more equitable and scalable retrieval systems.

\newpage
\bibliographystyle{ACM-Reference-Format}
\bibliography{multi_tower}

@inproceedings{zhou2018deep,
  title={Deep interest network for click-through rate prediction},
  author={Zhou, Guorui and Zhu, Xiaoqiang and Song, Chenru and Fan, Ying and Zhu, Han and Ma, Xiao and Yan, Yanghui and Jin, Junqi and Li, Han and Gai, Kun},
  booktitle={Proceedings of the 24th ACM SIGKDD international conference on knowledge discovery \& data mining},
  pages={1059--1068},
  year={2018}
}

@article{naumov2019deep,
  title={Deep learning recommendation model for personalization and recommendation systems},
  author={Naumov, Maxim and Mudigere, Dheevatsa and Shi, Hao-Jun Michael and Huang, Jianyu and Sundaraman, Narayanan and Park, Jongsoo and Wang, Xiaodong and Gupta, Udit and Wu, Carole-Jean and Azzolini, Alisson G and others},
  journal={arXiv preprint arXiv:1906.00091},
  year={2019}
}

@inproceedings{zhu2025rankmixer,
  title={Rankmixer: Scaling up ranking models in industrial recommenders},
  author={Zhu, Jie and Fan, Zhifang and Zhu, Xiaoxie and Jiang, Yuchen and Wang, Hangyu and Han, Xintian and Ding, Haoran and Wang, Xinmin and Zhao, Wenlin and Gong, Zhen and others},
  booktitle={Proceedings of the 34th ACM International Conference on Information and Knowledge Management},
  pages={6309--6316},
  year={2025}
}

@article{zhang2025onetrans,
  title={OneTrans: Unified Feature Interaction and Sequence Modeling with One Transformer in Industrial Recommender},
  author={Zhang, Zhaoqi and Pei, Haolei and Guo, Jun and Wang, Tianyu and Feng, Yufei and Sun, Hui and Liu, Shaowei and Sun, Aixin},
  journal={arXiv preprint arXiv:2510.26104},
  year={2025}
}

@inproceedings{wang2021dcn,
  title={Dcn v2: Improved deep \& cross network and practical lessons for web-scale learning to rank systems},
  author={Wang, Ruoxi and Shivanna, Rakesh and Cheng, Derek and Jain, Sagar and Lin, Dong and Hong, Lichan and Chi, Ed},
  booktitle={Proceedings of the web conference 2021},
  pages={1785--1797},
  year={2021}
}

@inproceedings{song2019autoint,
  title={Autoint: Automatic feature interaction learning via self-attentive neural networks},
  author={Song, Weiping and Shi, Chence and Xiao, Zhiping and Duan, Zhijian and Xu, Yewen and Zhang, Ming and Tang, Jian},
  booktitle={Proceedings of the 28th ACM international conference on information and knowledge management},
  pages={1161--1170},
  year={2019}
}

@article{zhang2022dhen,
  title={DHEN: A deep and hierarchical ensemble network for large-scale click-through rate prediction},
  author={Zhang, Buyun and Luo, Liang and Liu, Xi and Li, Jay and Chen, Zeliang and Zhang, Weilin and Wei, Xiaohan and Hao, Yuchen and Tsang, Michael and Wang, Wenjun and others},
  journal={arXiv preprint arXiv:2203.11014},
  year={2022}
}

@inproceedings{guo2019pal,
  title={PAL: a position-bias aware learning framework for CTR prediction in live recommender systems},
  author={Guo, Huifeng and Yu, Jinkai and Liu, Qing and Tang, Ruiming and Zhang, Yuzhou},
  booktitle={Proceedings of the 13th ACM Conference on Recommender Systems},
  pages={452--456},
  year={2019}
}

@inproceedings{lee2019melu,
  title={Melu: Meta-learned user preference estimator for cold-start recommendation},
  author={Lee, Hoyeop and Im, Jinbae and Jang, Seongwon and Cho, Hyunsouk and Chung, Sehee},
  booktitle={Proceedings of the 25th ACM SIGKDD international conference on knowledge discovery \& data mining},
  pages={1073--1082},
  year={2019}
}

@inproceedings{yi2019sampling,
  title={Sampling-bias-corrected neural modeling for large corpus item recommendations},
  author={Yi, Xinyang and Yang, Ji and Hong, Lichan and Cheng, Derek Zhiyuan and Heldt, Lukasz and Kumthekar, Aditee and Zhao, Zhe and Wei, Li and Chi, Ed},
  booktitle={Proceedings of the 13th ACM conference on recommender systems},
  pages={269--277},
  year={2019}
}

@inproceedings{chai2025longer,
  title={Longer: Scaling up long sequence modeling in industrial recommenders},
  author={Chai, Zheng and Ren, Qin and Xiao, Xijun and Yang, Huizhi and Han, Bo and Zhang, Sijun and Chen, Di and Lu, Hui and Zhao, Wenlin and Yu, Lele and others},
  booktitle={Proceedings of the Nineteenth ACM Conference on Recommender Systems},
  pages={247--256},
  year={2025}
}

@inproceedings{rendle2010factorization,
  title={Factorization machines},
  author={Rendle, Steffen},
  booktitle={2010 IEEE International conference on data mining},
  pages={995--1000},
  year={2010},
  organization={IEEE}
}

@inproceedings{juan2016field,
  title={Field-aware factorization machines for CTR prediction},
  author={Juan, Yuchin and Zhuang, Yong and Chin, Wei-Sheng and Lin, Chih-Jen},
  booktitle={Proceedings of the 10th ACM conference on recommender systems},
  pages={43--50},
  year={2016}
}

@article{zhang2024wukong,
  title={Wukong: Towards a scaling law for large-scale recommendation},
  author={Zhang, Buyun and Luo, Liang and Chen, Yuxin and Nie, Jade and Liu, Xi and Guo, Daifeng and Zhao, Yanli and Li, Shen and Hao, Yuchen and Yao, Yantao and others},
  journal={arXiv preprint arXiv:2403.02545},
  year={2024}
}

@inproceedings{wang2023fresh,
  title={Fresh Content Needs More Attention: Multi-funnel Fresh Content Recommendation},
  author={Wang, Jianling and Lu, Haokai and Zhang, Sai and Locanthi, Bart and Wang, Haoting and Greaves, Dylan and Lipshitz, Benjamin and Badam, Sriraj and Chi, Ed H and Goodrow, Cristos J and others},
  booktitle={Proceedings of the 29th ACM SIGKDD Conference on Knowledge Discovery and Data Mining},
  pages={5082--5091},
  year={2023}
}

% \appendix




\end{document}